\documentclass[twocolumn]{aastex62}

\turnoffediting

\graphicspath{{./}{figures/}}

\received{March 17, 2021}
\revised{April 7, 2021}
\submitjournal{ApJL}

\shorttitle{TRGB Distance to NGC\,1052--DF2}
\shortauthors{Shen et al.}

\begin{document}

\title{A Tip of the Red Giant Branch Distance of $22.1 \pm 1.2$ Mpc to the Dark Matter Deficient Galaxy NGC\,1052--DF2 from 40 Orbits of Hubble Space Telescope Imaging}


\correspondingauthor{Zili Shen}
\email{zili.shen@yale.edu}

\author[0000-0002-5120-1684]{Zili Shen}
\affiliation{Astronomy Department, Yale University,
52 Hillhouse Ave,
New Haven, CT 06511, USA}

\author[0000-0002-1841-2252]{Shany Danieli}
\altaffiliation{NASA Hubble Fellow}
\affiliation{Astronomy Department, Yale University,
52 Hillhouse Ave,
New Haven, CT 06511, USA}
\affiliation{Department of Physics, Yale University, New Haven, CT 06520, USA}
\affiliation{Yale Center for Astronomy and Astrophysics, Yale University, New Haven, CT 06511, USA}
\affiliation{Institute for Advanced Study, 1 Einstein Drive, Princeton, NJ 08540, USA}

\author[0000-0002-8282-9888]{Pieter van Dokkum}
\affiliation{Astronomy Department, Yale University,
52 Hillhouse Ave,
New Haven, CT 06511, USA}

\author[0000-0002-4542-921X]{Roberto Abraham}
\affiliation{David A. Dunlap Department of Astronomy and Astrophysics, University of Toronto, Toronto ON, M5S 3H4, Canada}
\affiliation{Dunlap Institute for Astronomy and Astrophysics, University of Toronto, Toronto ON, M5S 3H4, Canada}

\author[0000-0002-9658-8763]{Jean P. Brodie}
\affiliation{University of California Observatories, 1156 High Street, Santa Cruz, CA 95064, USA}

\author[0000-0002-1590-8551]{Charlie Conroy}
\affiliation{Harvard-Smithsonian Center for Astrophysics, 60 Garden Street, Cambridge, MA, USA}

\author[0000-0001-8416-4093]{Andrew E. Dolphin}
\affiliation{Raytheon Technologies, 1151 E. Hermans Road, Tucson, AZ 85756, USA}
\affiliation{Steward Observatory, University of Arizona, 933 North Cherry Avenue, Tucson, AZ 85721-0065 USA}

\author[0000-0003-2473-0369]{Aaron J. Romanowsky}
\affiliation{Department of Physics and Astronomy, San Jose State University, San Jose, CA 95192, USA}
\affiliation{University of California Observatories, 1156 High Street, Santa Cruz, CA 95064, USA}

\author[0000-0002-8804-0212]{J.~M.~Diederik Kruijssen}
\affiliation{Astronomisches Rechen-Institut, Zentrum f\"ur Astronomie der Universit\"at Heidelberg, M\"onchhofstraße 12-14, D-69120 Heidelberg, Germany}

\author[0000-0003-0250-3827]{Dhruba Dutta Chowdhury}
\affiliation{Astronomy Department, Yale University,
52 Hillhouse Ave,
New Haven, CT 06511, USA}

\begin{abstract}

The large and diffuse galaxies NGC\,1052--DF2 and NGC\,1052--DF4 have been found to have very low dark matter content and a population of luminous globular clusters. Accurate distance measurements are key to interpreting these observations. Recently, the distance to NGC\,1052--DF4 was found to be $20.0\pm 1.6$\,Mpc by identifying the tip of the red giant branch (TRGB) in 12 orbits of \textit{Hubble Space Telescope (HST)} Advanced Camera for Surveys (ACS) imaging. Here we present 40 orbits of \textit{HST} ACS data for NGC\,1052--DF2 and use these data to measure its TRGB.
The TRGB is readily apparent in the color-magnitude diagram. Using a forward model that incorporates photometric uncertainties, we find a TRGB magnitude of $m_{\rm F814W, TRGB} = 27.67 \pm 0.10$\,mag. The inferred distance is $D_{\rm TRGB} = 22.1 \pm 1.2$ Mpc, consistent with the \edit1{previous surface brightness fluctuation distances} to the bright elliptical galaxy NGC\,1052. The new \textit{HST} distance 
rules out the idea that some of NGC\,1052--DF2's unusual properties can be explained if it were at $\sim 13$\,Mpc; instead, it implies that the galaxy's globular clusters are even more luminous than had been derived using the previous distance \edit1{of 20\,Mpc}. The distance from NGC\,1052--DF2 to NGC\,1052--DF4 is well-determined at $2.1\pm 0.5$\,Mpc, significantly larger than the virial diameter of NGC\,1052. We discuss the implications for formation scenarios of the galaxies and for the external field effect, which has been invoked to explain the intrinsic dynamics of these objects in the context of modified Newtonian dynamics. 
\end{abstract}


\section{Introduction} \label{sec:intro}
First identified by \citet{Karachentsev2000}, NGC\,1052--DF2 is a galaxy in one of the fields of the Dragonfly Nearby Galaxies Survey \citep{Merritt2016} that was found to be unusual in follow-up  observations \citep{vanDokkum2018Nature,Cohen2018}. At the distance of the bright elliptical galaxy NGC\,1052 \citep[D = $19.4-21.4$\,Mpc;][]{Blakeslee2001,2001ApJ...546..681T,Tully2013}, NGC\,1052--DF2 has the stellar mass ($\sim 2 \times 10^8 M_{\odot}$) and metallicity of typical dwarf galaxies \citep{Fensch2019} but a large size ($R_{\rm e} = 2.2$ kpc), placing it in the regime of ultra-diffuse galaxies \citep{vanDokkum2015}.
Furthermore, it has a large population of luminous globular clusters (GCs) and a very low velocity dispersion of $8.5^{+2.3}_{-3.1}$\,km\,s$^{-1}$ \citep[see][]{vanDokkum2018Nature,Danieli2019,Emsellem2019,Shen2020}. In 2019, a second galaxy in the same field, NGC\,1052--DF4, was found to have properties almost identical to those of NGC\,1052--DF2 \citep{VanDokkum2019DF4,Danieli2020,Shen2020}.
The velocity dispersions of both galaxies 
\citep[$4-10$\,km\,s$^{-1}$;][]{vanDokkum2018Nature,VanDokkum2019DF4,Danieli2019,Emsellem2019}
are consistent with expectations from the stellar mass alone ($\approx 7$\,km\,s$^{-1}$), suggesting that the galaxies have no or very little dark matter \citep{vanDokkum2018Nature}.
\citet{vanDokkum2018Nature} claimed that NGC\,1052--DF2 invalidates alternative gravity theories such as Modified Newtonian Dynamics \citep[MOND;][]{Milgrom1983}, as the (apparent) effects of dark matter should be observed in all galaxies in these models.

The unusual nature of NGC\,1052--DF2 and NGC\,1052--DF4 sparked a vigorous debate in the community. 
Whereas initial concerns focused on the velocity dispersion measurements \citep[see, e.g., ][]{Martin2018}, more recently the distance to the galaxies has become a point of contention. 
Their radial velocities of $1803$\,km\,s$^{-1}$ and $1445$\,km\,s$^{-1}$, their projected proximity to the elliptical galaxy NGC\,1052, and surface brightness fluctuations in 1+1 orbit F814W and F606W \textit{HST} images all indicate a distance of $\approx 20$\,Mpc for the galaxies \citep{vanDokkum2018Nature,Blakeslee2018,Cohen2018,vanDokkum2018dist}.  However,  \citet{Trujillo2019} and \citet{Monelli2019} derived distances of 13--14\,Mpc to both NGC\,1052--DF2 and NGC\,1052--DF4 from the same 1+1 orbit \textit{HST} images, claiming
that individual red giant stars were detected and associating a sharp increase in the number of detections around $m_{F814W} \approx26.5$ mag to be the tip of the red giant branch (TRGB). A distance of 13\,Mpc is one way to resolve or alleviate the unusual properties of NGC\,1052--DF2 and NGC\,1052--DF4: for a smaller distance, the galaxies are no longer ``ultra-diffuse'', the GCs are smaller and less luminous, and the dark matter fraction increases.

Accurate distances to both galaxies are important for deriving accurate luminosities, sizes, and masses -- of both the galaxies themselves and their GCs -- and are also key to placing constraints on alternative gravity models such as MOND. As pointed out by \citet{Kroupa2018}, \citet{Muller2019MOND}, \citet{Haghi2019}, and \citet{Famaey2018},  \citet{vanDokkum2018Nature} had neglected an essential element of MOND, namely the external field effect (EFE). 
The EFE, unique to MOND, causes a low mass galaxy in orbit around a massive galaxy to have a lower velocity dispersion than the same object in isolation.
Thus, the low velocity dispersions of NGC\,1052--DF2 and NGC\,1052--DF4  may be consistent with the expectations from MOND if both galaxies are in
close proximity to NGC\,1052 \citep{Kroupa2018}. 

Accurate distances can be obtained using \textit{HST} data that are much deeper than the 1+1 orbits in the \citet{Cohen2018} study, as the TRGB measurement is sensitive to the depth and quality cuts used in the analysis. In a single F814W orbit, individual red giants can only be detected out to $\approx 14$\,Mpc, right where \citet{Trujillo2019} and \citet{Monelli2019} claimed to see the onset of the giant branch. 
With $\approx 10$ orbits, the TRGB can be detected out to 20\,Mpc.
\citet{Danieli2020} obtained the needed data for NGC\,1052--DF4, and using a total of 8 orbits in F814W and 4 orbits in F606W measured a TRGB distance to that galaxy of $20.0 \pm 1.6$ Mpc. 
Here, we present even deeper \textit{HST} imaging (20+20 orbits in F814W and F606W) for NGC\,1052--DF2.
We unambiguously identify the TRGB and derive a distance to the galaxy that should be definitive. Furthermore, we derive an accurate and (as it turns out) interesting {\em relative} distance between NGC\,1052--DF2 and NGC\,1052--DF4, making use of the fact that nearly all systematic uncertainties cancel in the comparison. This relative distance provides important context for the presence of tidal features in the outskirts of the galaxies \citep{Montes2020} and constrains the EFE.
Vega magnitudes are used throughout this paper.

\section{\textit{HST}/ACS Data} \label{sec:HST}
\begin{figure*}
\plotone{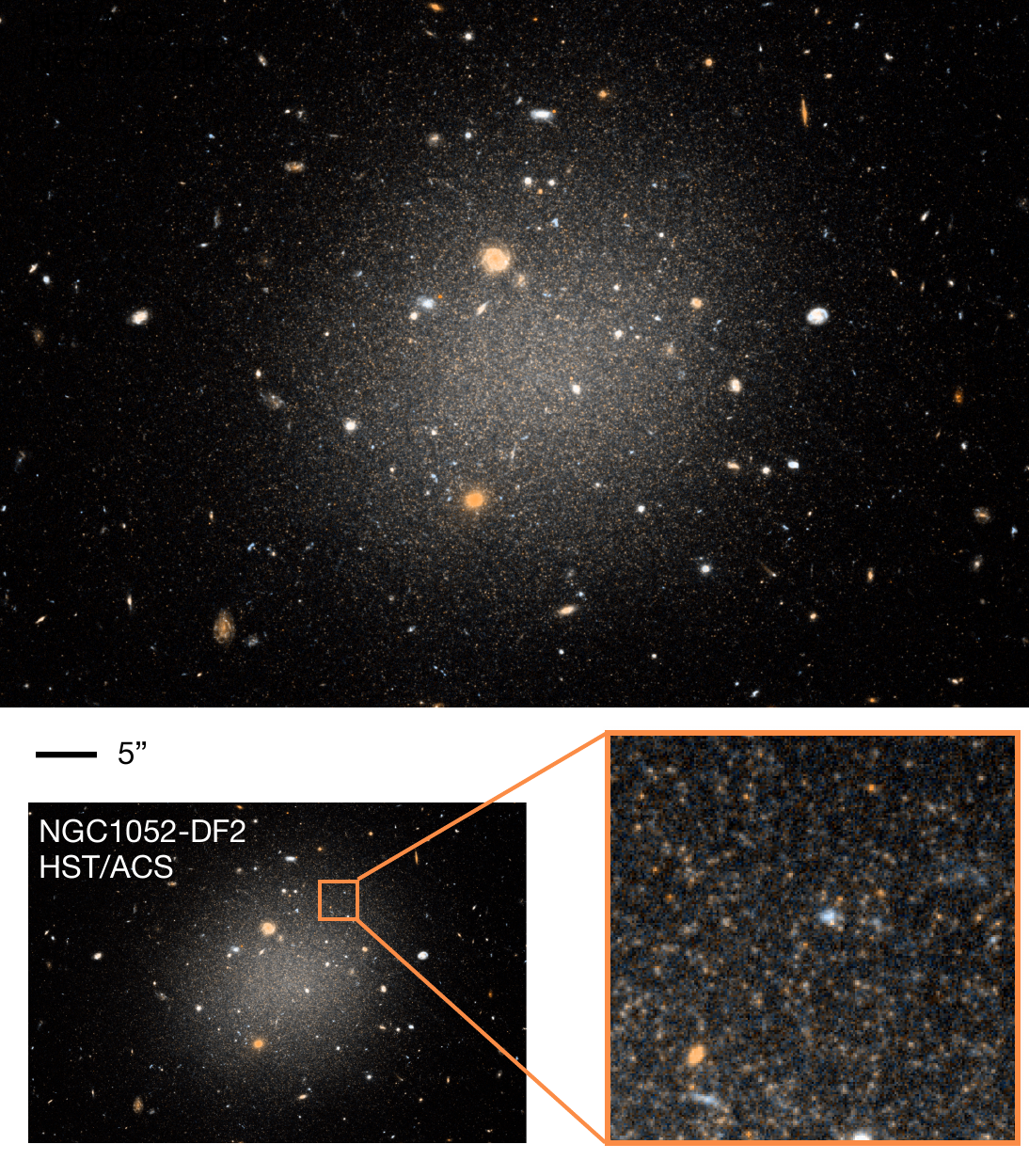}
\caption{Deep image of NGC\,1052--DF2 obtained with \textit{HST}/ACS.
The total exposure time is 41840s in F814W and 40560s in F606W, 20 orbits each.
\edit2{North is up and east is to the left. The scale bar is for the upper panel ($\sim0.5$ kpc at the distance obtained in this work).} The \edit1{bottom panel} shows \edit1{a zoom and reveals} resolved red giant stars in a background of unresolved blue stars.}
\label{fig:img}
\end{figure*}

NGC\,1052--DF2 was observed with \textit{HST} ACS in Cycle 27 (program 15851), split in three epochs from June 2020 to September 2020. We obtained 19 orbits each in the F814W and F606W filters. Adding these data to the previous 1 orbit each in the same filters obtained in 2017 (program 14644), the total exposure time is 41840s in F814W and 40560s in F606W. NGC\,1052--DF2 was placed near the center of the field of view, varying the exact location between visits to ensure relatively uniform coverage despite the chip gap. 

Within each orbit, four exposures were taken in a four-point dither pattern to remove cosmic rays, identify hot pixels, and dither over the chip gap. The STScI standard pipeline was used to perform bias and dark current subtraction, flat-fielding, and CTE correction on individual exposures and produces calibrated \texttt{flc} files. We used the \texttt{TweakReg} utility in DrizzlePac to align the 160  \texttt{flc} files. We used AstroDrizzle to remove astrometric distortion, correct sky background, and flag cosmic-rays in the aligned images, resulting in a combined (\texttt{drc}) image in the F814W and F606W filters.

Figure \ref{fig:img} shows a color image of NGC\,1052--DF2 generated from the drizzled images. In these deep data, the galaxy is well-resolved. The yellow points are resolved red giants which appear among a background of bluer unresolved subgiants and main sequence stars. We do not confirm the ``Maybe Stream'' proposed in \cite{Abraham2018}. The clumps of luminous giants that make up the apparent feature are detected but we do not find an accompanying excess of fainter stars that would be present if it were a tidal feature. We conclude that the apparent feature is a chance alignment of background galaxies and Poisson fluctuations in the number of bright stars.

\section{Photometry} \label{sec:phot}
The software package DOLPHOT\footnote{http://americano.dolphinsim.com/dolphot/} \citep{2000PASP..112.1383D} was used to carry out photometry on individual \texttt{flc} files from the STScI ACS pipeline.
Before running the photometry detection step, 
all 160 \texttt{flc} files as well as the deepest drizzled image (which combines 80 stacked F814W  \texttt{flc} images with a total exposure time of 41840s)
were ran through several preparation steps using the DOLPHOT/ACS pre-processing tools. We used the \texttt{acsmask} routine to mask bad columns and hot pixels, the \texttt{splitgroups} routine to split each \texttt{flc} into two chips, and the \texttt{calcsky} routine to determine the background value in each image.

DOLPHOT analyzed all 160 \texttt{flc} images simultaneously, using the deep combined \edit1{F814W} \texttt{drc} image as reference. 
Stars were detected in each image by fitting Tiny Tim PSFs \citep{1995ASPC...77..349K} and 
the fluxes were measured separately in the F814W and F606W frames. 
Although this approach is memory-intensive for a deep data set like ours, the fact that no resampling is required has the advantages of preserving the noise properties and optimally using the information in the \texttt{flc} files.
The key DOLPHOT parameters were: the sky fitting parameter \texttt{FitSky}=2,
the aperture radius \texttt{RAper}=3, and \texttt{Force1}=1 which forces the sources to be fitted as stars. 
These parameters were identical to the values used for the TRGB analysis on NGC\,1052--DF4 \citep{Danieli2020}, and we refer to that paper for further information.

Stars were selected from the raw DOLPHOT output catalog with a set of quality cuts.
The photometry catalog was corrected for Galactic extinction: 0.04 mag in F814W and 0.06 mag in F606W \citep{2011ApJ...737..103S}.
Objects were considered to be reliable stars only if they pass all of the following criteria: 
signal-to-noise ratio $>4$ in F814W and $>3$ in F606W; 
object-type $\leq 2$, which corresponds to \textit{good star} or \textit{faint star}); 
sharpness parameter $|\text{sharp}_{F814W}| \leq 0.5$ and $|\text{sharp}_{F606W}| \leq 0.5$;
crowding parameter $|\text{crowd}_{F814W}| \leq 0.5$ and $|\text{crowd}_{F606W}| \leq 0.5$.
Compared to the criteria  in \citet{Danieli2020}, we increased the signal-to-noise threshold in F606W to match our deeper data.

Characterization of the systematic and random uncertainties in the photometry is an important component of the analysis, as they have an impact on the apparent location of the TRGB \citep[see, e.g.][]{2006AJ....132.2729M}.
In DOLPHOT, stellar PSFs (using models appropriate for their individual $x,y$ positions) can be placed in the \textit{HST} data, and then analyzed alongside the actual stars. The scatter and systematic offset in the magnitudes of injected and recovered stars can then be taken into account in the measurement of the TRGB location.
We ran DOLPHOT photometry on 180,000 artificial stars. 
The stars uniformly sampled the magnitude range $25< F814W < 29$ and the color range $1.0<F606W-F814W<1.5$, and were distributed between $R_{\text{eff}}$ $(\approx 20\arcsec)$ and $4\,R_{\text{eff}}$.
The artificial stars were injected into individual \texttt{flc} images and analyzed with the same methodology as the real stars.

\begin{figure*}
    \centering
    \includegraphics[width=\textwidth]{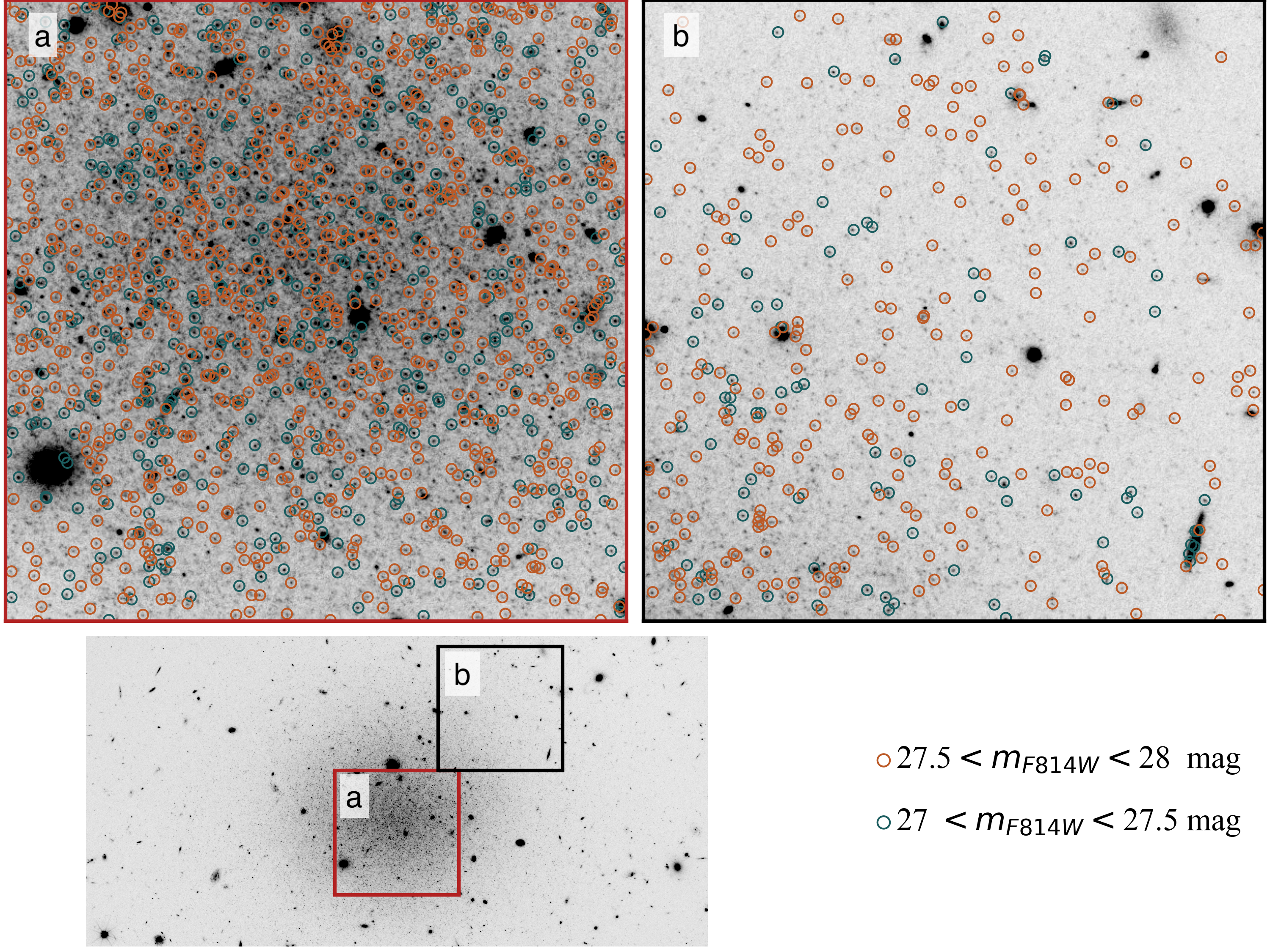}
    \caption{Stars detected after quality cuts in NGC\,1052--DF2, shown in small regions in the \edit1{center (panel a}) and \edit1{the outskirts (panel b}).
    \edit1{The TRGB magnitude is approximately 27.5 mag, as identified in \S\,\ref{subsec:edge}.}
    Stars with $27.5< m_{\text{F814W}} < 28.0$ are circled in orange and stars  with $27.0< m_{\text{F814W}} < 27.5$ in green. The number of sources brighter than 27.5 mag decrease visibly on the image. Blended objects (especially in the center) do not pass the quality cuts. Stars in the central part of the galaxy ($R< R_{\text{eff}}$) are excluded from the analysis.}
    \label{fig:phot}
\end{figure*}

Figure \ref{fig:phot} shows examples of stellar detections in two regions of NGC\,1052--DF2. 
There is a sharp increase in the number of sources fainter than 27.5 mag, \edit1{which we identify as the approximate location of the TRGB in \S\,\ref{subsec:edge} with edge detection and \S\,\ref{subsec:modeling} with forward modeling. We find} $\sim 3$ times more sources detected with $27.5<m_{\rm F814W}<28.0$ than with $27.0<m_{\rm F814W}<27.5$.
\edit1{The sudden increase at 27.5 mag is especially visible in the outer regions of the galaxy, where the stellar density is lower and crowding is less severe.}\deleted{In the following section, this result is quantified using a simple edge detection, followed by a fit of the TRGB magnitude.}


\section{TRGB Distance} \label{sec:dist}
\subsection{Color-magnitude Diagrams} \label{subsec:CMD}

\begin{figure*}
\plotone{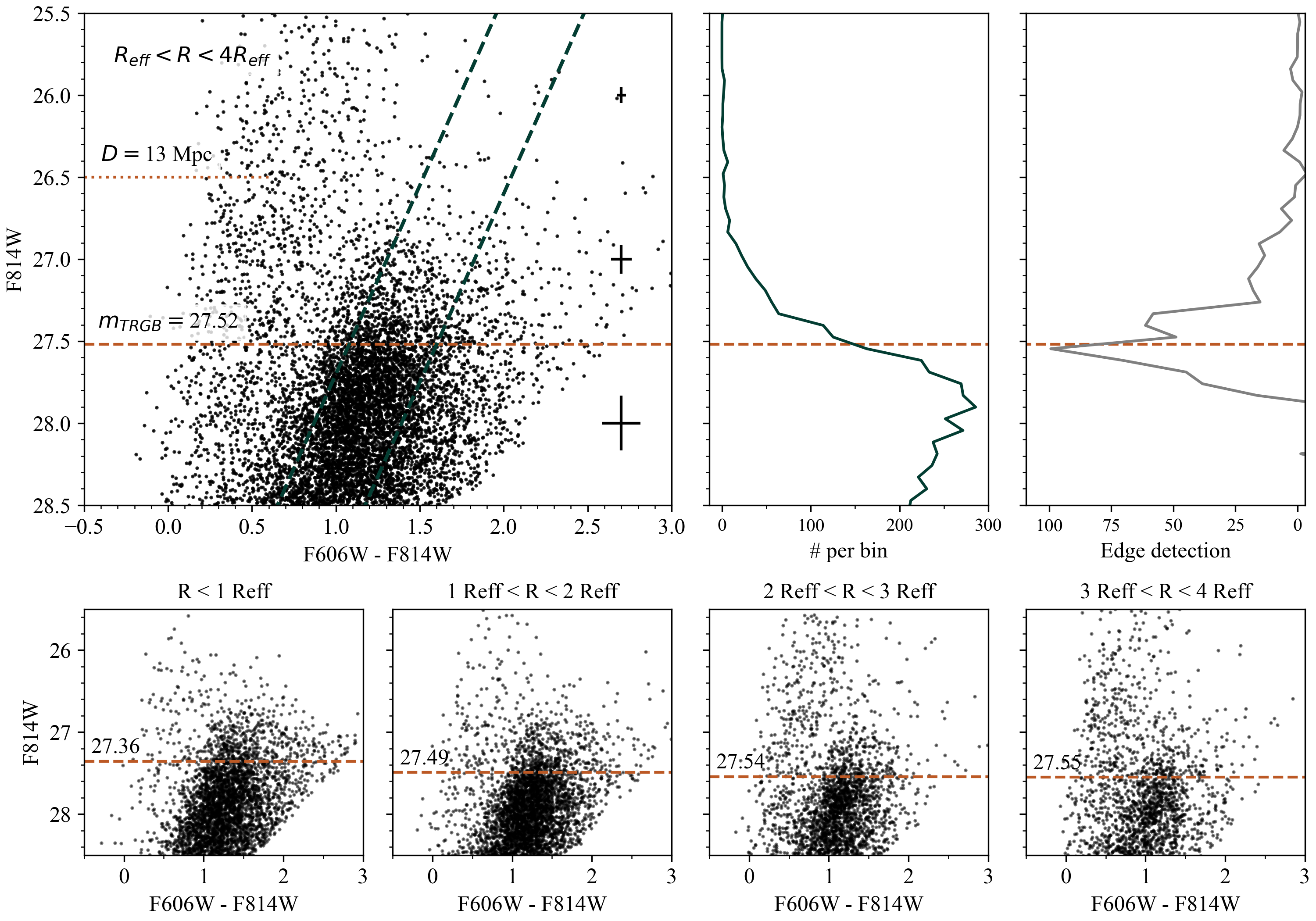}
\caption{TRGB edge detection for NGC\,1052--DF2.
Top panels: the color--magnitude diagram (left), the binned and smoothed F814W luminosity function (middle), and its response function to a sobel kernel (right) are shown for stars located at $R_{\text{eff}} < R < 4R_{\text{eff}}$. Black points are stars that pass quality cuts, and the luminosity function is calculated for stars that pass an additional color cut to select the RGB (the diagonal dashes lines). Crosses indicate photometric scatter. 
\deleted{The dashed orange line marks the location of the TRGB }
\edit1{The dashed orange line indicates the steep increase in the F814W luminosity function that we identify as the TRGB,}
measured by fitting a Gaussian, $m_{\text{TRGB,F814W}} = 27.52 \pm 0.17$ mag, while the dotted orange line marks the TRGB for 13\,Mpc \citep{Trujillo2019}.
Bottom panels: the color-magnitude diagrams of stars in radial bins. The dashed line indicates the respective TRGB measurement.}
\label{fig:CMD}
\end{figure*}

Color--magnitude diagrams (CMDs) were constructed for different radial regions in NGC\,1052--DF2 and shown in Figure \ref{fig:CMD}.
The main panel and the bottom panels of Figure \ref{fig:CMD} all show the characteristic sharp decrease in the number of stars around $m_{F814W} \sim 27.5$ mag, indicating the onset of the red giant branch (RGB)\footnote{The photometry catalog is available at \url{https://github.com/zilishen/NGC1052-DF2-public}.}.
The F814W luminosity\edit1{, with bin width 0.07 mag, } is shown in the middle panel, as measured within the following color range:
$2< (F814W - 28) + 2.3\times(F606W-F814W) < 3.2$. This luminosity function is fitted in \S\,\ref{subsec:edge} and \S\,\ref{subsec:modeling} to identify
the location of the TRGB, which is then used to derive the distance.

\subsection{Edge Detection} \label{subsec:edge}
The first method, edge detection, measures the TRGB from the first--derivative of the binned and smoothed luminosity function of the RGB and asymptotic giant branch (AGB) stars.
This method was previously used for NGC\,1052--DF4 \citep{Danieli2020} and numerous other galaxies 
\citep[e.g.,][]{1993ApJ...417..553L}. Advantages of the method include that it is straightforward to compare results to other studies and other galaxies, and that it provides a robust measurement of the approximate location of the TRGB. We caution, however, that it does not take  measurement uncertainties into account.\footnote{In practice, the systematic bias (see \S\,\ref{subsec:modeling}) causes the TRGB to appear fainter than the true value, whereas photometric scatter causes it to appear brighter.}
The $F814W$-band luminosity function was smoothed with a Gaussian kernel with
a standard deviation of 0.4 mag and then filtered with a Sobel kernel. 
The response to the Sobel kernel peaks where the slope of the luminosity function reaches a maximum, and this is the location of the TRGB.
The peak location was identified by performing a Gaussian fit to the response function to determine its center and width.

The TRGB measurement we obtained with the edge-detection method is shown in Figure \ref{fig:CMD}. 
The main result came from a sample in the radial region $R_{\text{eff}} < R < 4R_{\text{eff}}$ (upper left panel).
We measure an extinction--corrected $F814W$ TRGB magnitude of $m_{\text{TRGB,F814W}} = 27.51 \pm 0.22$ mag for NGC\,1052--DF2, where the uncertainty is the width of the Gaussian fit to the peak.
The CMD in each radial bin from $R_{\text{eff}}$ to $4R_{\text{eff}}$ and the corresponding measured TRGB magnitude is shown in the bottom row of Figure \ref{fig:CMD}.
The measurement is consistent within error bars across all radial bins, even in the innermost region where crowding is most severe and many initial source detections are rejected by the quality cuts. We infer that the TRGB is located near $m_{\rm F814W}\sim 27.5$, and not near $m_{\rm F814W}\sim 26.5$ as was found by \citet{Trujillo2019}. 

\subsection{Forward Modeling} \label{subsec:modeling}

\begin{figure*}
\plotone{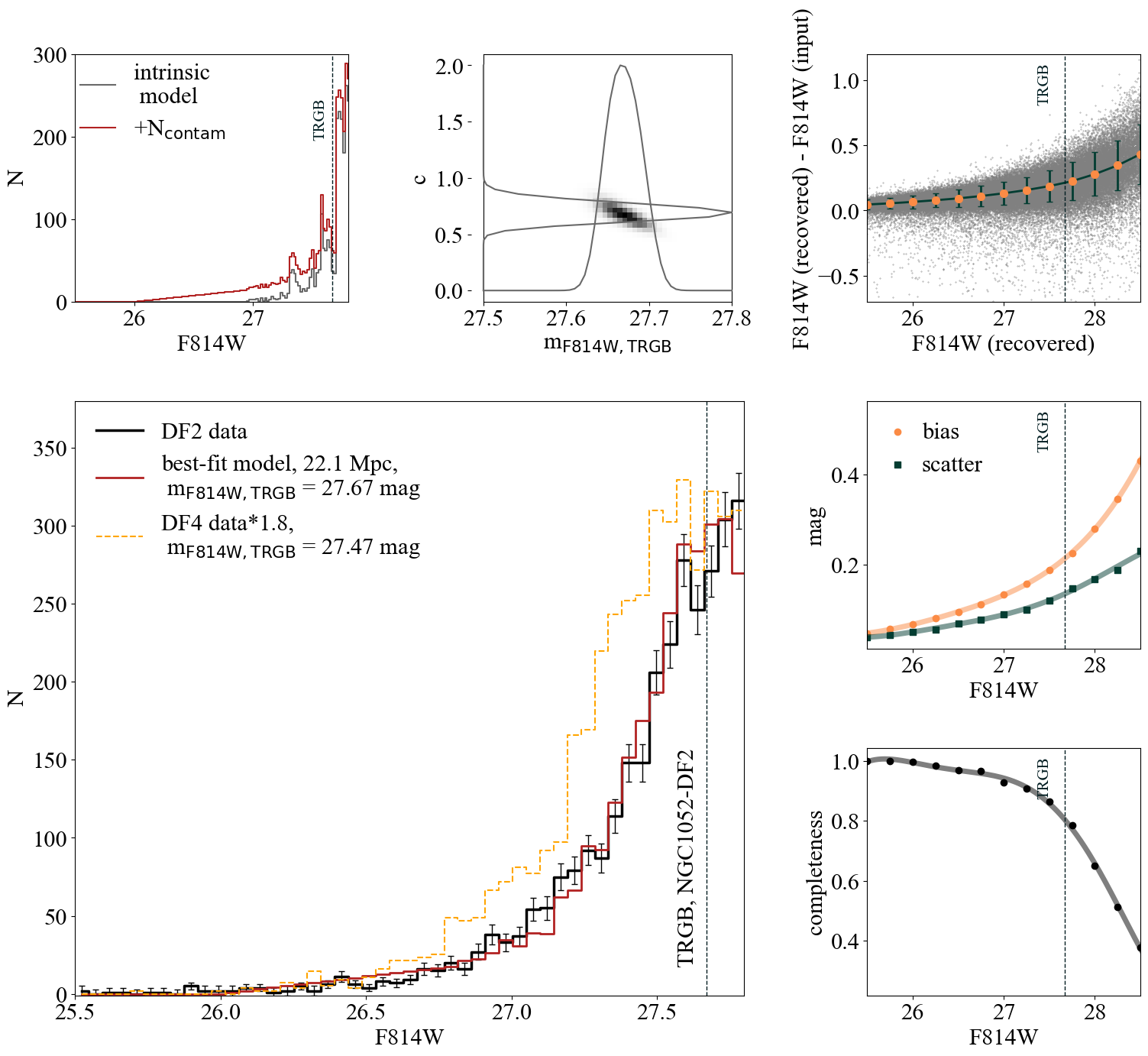}
\caption{TRGB forward modeling results for NGC\,1052--DF2.
Top \edit1{left}: the F814W luminosity function of the intrinsic model  \edit1{without contamination (grey) and with contamination added (red).}\deleted{before applying photometric errors (left) and the likelihood (middle).}
\edit1{Top middle: the likelihood map from 1000 Monte Carlo runs, marginalized over the TRGB magnitude $m_{\text{TRGB,F814W}}$ and the contamination parameter $c$, as defined in Eq. \ref{eq:contam}. The parameter values at maximum likelihood are used to construct the best-fit model in the main panel.}
\deleted{Right panels show the}\edit1{Top right: the} systematic offset (bias\edit1{, in orange}) and the scatter \edit1{(in green)} in the recovered magnitudes using the artificial star test\edit1{. Middle right: the bias and scatter from top right panel, fitted with a fifth-order polynomial. Bottom right: photometric completeness, the fraction of input artificial stars that are recovered at each magnitude.}\deleted{(top and middle), and the completeness (bottom).} 
Main panel: the best-fit model (red) of the observed F814W luminosity function (black), with a TRGB magnitude of $m_{\text{TRGB,F814W}} = 27.67$
\edit1{from the maximum likelihood}.
\edit1{The best-fit model is the intrinsic luminosity function with contamination added (shown in the top left panel) and photometric errors applied (shown in the right panels).}
The reduced $\chi^2$ value for this fit is 1.8.
For comparison, the observed luminosity function of NGC\,1052--DF4 \citep{Danieli2020} is scaled and plotted in the dashed yellow line.
}
\label{fig:forward-model}
\end{figure*}

The second method to measure the TRGB is fitting the \deleted{I-band}\edit1{F814W} luminosity functions. 
We derived a  precise TRGB location with the same forward modeling procedure as \citet{Danieli2020}.
Briefly summarized, this approach generates an intrinsic luminosity function of a stellar population by drawing stars from an isochrone according to a initial mass function, then perturbs the photometry of individual stars according to the (magnitude-dependent) photometric errors to obtain an observed luminosity function, and finally adds contaminating stars according to a simple linear distribution in magnitude. As the \deleted{location}\edit1{luminosity} of the \deleted{giant branch}\edit1{TRGB} is not very sensitive to age or metallicity, the results are not sensitive to the details of the stellar population synthesis model; following \citet{Danieli2020} we used an old (10 Gyr),  metal-poor ([Fe/H] $=-1$) stellar population.
\edit2{The population is not limited to RGB stars but includes stars in other phases, particularly the AGB.}
\edit2{This stellar population is chosen to reproduce the observed color and mass of the galaxy, and informed by previous spectroscopic studies: both the diffuse light
\citep{Fensch2019} and the GCs \citep{vanDokkum2018Nature} are old and low-metallicity.}

The two \deleted{free}\edit1{fit} parameters, the TRGB magnitude and the contamination, were varied to search for a best fit to the F814W luminosity function.

The photometric errors were characterized using artificial star tests (see \S\,\ref{sec:phot}). Compared to the input magnitudes of injected artificial stars, the recovered magnitudes show a systematic offset (``bias''), scatter, and incompleteness. 
\edit1{After applying the same quality cuts listed in \S\,\ref{sec:phot} to the artificial stars, we calculated the difference between recovered and input magnitudes. In each magnitude bin of width 0.25 mag, the median value of the difference was taken to be the bias, the standard deviation was the scatter, and the fraction of recovered input stars was the completeness. Each of these quantities was fitted by a fifth-order polynomial in the range 25.5 mag to 29.0 mag.}
These results are shown in the right panels of Figure \ref{fig:forward-model}. 
The bias \edit1{(orange data points and curves)} and scatter \edit1{(green data points and curves)} increase towards fainter magnitudes and completeness decreases as expected for fainter stars.
At $m_{\text{F814W}} = 27.67$, the recovered magnitudes are 0.21 mag fainter than the input magnitudes, with a $1\sigma$ scatter of 0.13 mag, and the completeness fraction is 80\%.
Compared to the photometric errors of \citet{Danieli2020}, our deeper data show a similar level of bias and smaller scatter. 
We find a $\sim 0.5$ mag fainter 50\%-completeness limit, as expected from the increase from 8 orbits to 20 orbits in the F814W filter.

Starting from $10^9$ stars drawn from a MIST isochrone \citep{Dotter2016,Choi2016}, the magnitude of each star was shifted by the parameterized bias and perturbed by a random sample of the scatter.
Possible contamination was parametrized by a linear function 
\begin{equation}
    N_{\rm contam} = 20c \times (m-26)
    \label{eq:contam}
\end{equation}
for stars fainter than 26 mag, and no contamination for stars brighter than 26 mag.
\edit1{$N_{\rm contam}$ is the number of contaminants at magnitude $m$, and $c$ is a fit parameter.
This was the simplest contamination model that captures the data (the black histogram in Figure \ref{fig:forward-model}).}
After adding the contamination, the luminosity function was multiplied by the completeness fraction.

This model was fitted to the stars that satisfy the quality and color cuts and
are located between $R_{\text{eff}}$ and $4\,R_{\text{eff}}$ from the center.
The observed F814W luminosity function had 50 \edit1{equal-width} magnitude bins from 25.50 to 27.85 mag, and the best fit was found by minimizing $\chi^2$ while varying  $m_{TRGB}$ and $c$. 
\edit1{We ran 1000 Monte Carlo fits, where the fitting range is $27.5<m_{\text{F814W}} < 27.8$ mag and $0<c<2$. The resulting likelihood map and the marginalized likelihood distributions are shown in the top middle panel of Figure \ref{fig:forward-model}.}
We obtained a best-fit TRGB magnitude of  $27.67 \pm 0.02$, where
\edit1{the marginal likelihood is maximized and}
the error is
the statistical uncertainty that encompasses 68\% of the marginal likelihood.
The reduced $\chi^2$ of the best fit was 1.8.
The best-fit model is shown in the main panel of Figure \ref{fig:forward-model}. The fit is excellent, as it reproduces not only the overall increase in the number of stars at faint magnitudes but also several of the individual small ``steps'' in the luminosity function. To our knowledge, our previous paper on NGC\,1052--DF4 was the first application of this method; the fact that it is able to fit small scale features in the luminosity function may warrant wider application of this technique. 

The TRGB magnitude derived with forward modeling is fainter than that determined from edge detection.
This difference of $0.15$ mag, previously seen in \citet{Danieli2020}, is due to the effect of photometric scatter on the apparent location of the TRGB. Many more stars scatter from faint to bright magnitudes than the other way, due to the steep increase in the luminosity function around the TRGB. We adopt the forward modeling value as the best measurement\deleted{ but add 0.1\,mag systematic uncertainty}. \edit2{We estimate the systematic error in the measurement by considering the variation in the bias, scatter, completeness within our $1- 4\ R_{\text{eff}}$ range. From the inner region ($1- 2\ R_{\text{eff}}$) to the outer region ($3- 4\ R_{\text{eff}}$), the bias decreases by 0.03 mag, the scatter decreases by 0.07 mag, and the completeness increases by 0.07 mag. Adding these in quadrature, we find 0.1 mag systematic uncertainty.} The final result is therefore
$m_{\rm TRGB} = 27.67 \pm 0.02$ (random) $\pm 0.10$ (systematic).

From the TRGB measurement, we derive the distance to NGC\,1052--DF2. 
The zero-point calibration of the absolute magnitude of the TRGB depends on the color \citep{2007ApJ...661..815R}:
\begin{equation}
    M_{F814W}^{ACS} = -4.06 +0.20[(F606W - F814W) -1.23]
\end{equation}
For a TRGB median color of 1.25 mag,  as measured from the color-magnitude diagram,
the absolute magnitude of the TRGB is $M_{F814W} = -4.056$ mag, with a standard \deleted{systematic}\edit1{calibration} uncertainty of 0.07 mag \citep{2017AJ....154...51M}.

The errors on the final distance measurement have three components: the statistical uncertainty from the fitting procedure (0.02 mag), the systematic uncertainty from the artificial star tests (bias of 0.1 mag), and finally the zero-point calibration uncertainty of 0.07 mag. Adding these in quadrature leads to a final distance modulus of $31.72 \pm  0.12$ mag and a distance of $D = 22.1 \pm 0.2$ (statistical) $\pm 1.0$ (systematic) $\pm0.7$ (calibration) $= 22.1 \pm 1.2$ Mpc.

\section{Discussion} \label{sec:discussion}

\begin{figure*}
    \centering
    \includegraphics[width=\textwidth]{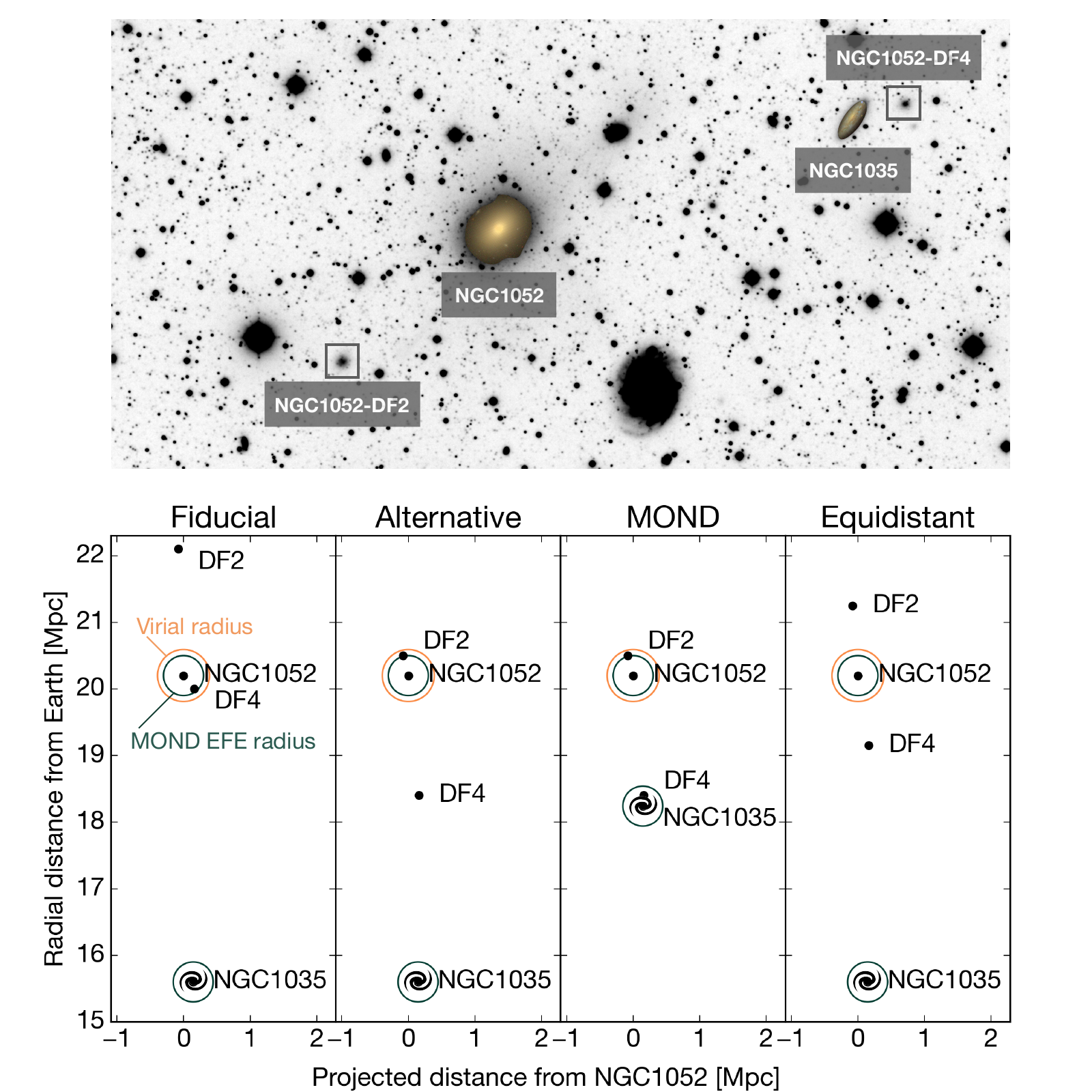}
    \caption{Top panel: Dragonfly r-band image of the NGC\,1052 field.
    Bottom panels: Four possible arrangements of NGC\,1052--DF2, NGC\,1052--DF4, NGC\,1052, and NGC\,1035. The fiducial panel shows the measured radial and projected distances of the four galaxies. The radial distances of NGC\,1052--DF2 and NGC\,1052--DF4 in other panels move within 1.5$\sigma$ error. The second panel shows that NGC\,1052--DF2 could be within range of EFE in MOND, but NGC\,1052--DF4 would be outside due to their relative distance of $2.1$ Mpc. The third panel is most consistent with MOND, where NGC\,1052--DF2 and NGC\,1052--DF4 are each within the EFE radius of a massive galaxy, but NGC\,1035 is 1.3$\sigma$ away from its fiducial distance. The rightmost panel shows a scenario where NGC\,1052--DF2 and NGC\,1052--DF4 are equidistant from NGC\,1052, possibly resulting from a high-speed encounter.}
    \label{fig:discussion}
\end{figure*}

In this study, we used extremely deep \textit{HST}/ACS data to measure the TRGB distance to NGC\,1052--DF2. Our best fit TRGB
distance is $22.1 \pm 1.2$\,Mpc. This distance to NGC\,1052--DF2 is just consistent with, but slightly larger than, two previous independent measurements of surface brightness fluctuations (SBF): $D_{\rm SBF}=18.7 \pm 1.7$ Mpc \citep{vanDokkum2018dist} and $D_{\rm SBF}=20.4 \pm 2.0$ Mpc \citep{Blakeslee2018}. The TRGB distance should be more reliable than the
SBF distances, as SBF absolute magnitudes are more model-dependent \citep[see, e.g.,][]{Greco2021}.
The new distance measurement to NGC\,1052--DF2 is consistent with the \edit1{SBF} distance to the elliptical galaxy NGC\,1052 within the uncertainties \citep[19.4 – 21.4 Mpc; ][]{Blakeslee2001,2001ApJ...546..681T,Tully2013}, \deleted{pressing towards the high end of the range}\edit1{but slightly larger than the value quoted in \citep{vanDokkum2018dist}}. The new \textit{HST} data rule out a distance to the galaxy of 13.6\,Mpc, which had been proposed by \citet{Trujillo2019}. The original single orbit \textit{HST} data were not deep enough to identify individual RGB stars at a distance of 20 Mpc, and as shown in \citet{vanDokkum2018dist}, the apparent feature at $m_{\rm F814W}\approx 26.5$ in those data can be reproduced by a combination of blends and AGB stars.

Our new distance confirms the unusual dark matter and GC properties of NGC\,1052--DF2, and
the slight upward revision of the distance to NGC\,1052--DF2 makes its luminous GCs even more spectacular.
The previously reported luminosities of the GCs in \citet{VanDokkum2018DF2GC} and \citet{Shen2020} assumed a distance of 20 Mpc. With the revised distance, all of the GCs in NGC\,1052--DF2 are 0.2 mag more luminous than previously reported. 
The GC luminosity function now peaks at $M_v \approx -9.3$, even further from the canonical value of $-7.5$ \citep[see][]{Rejkuba2012}.

Interestingly, the TRGB distance to NGC\,1052--DF2 is 2.1 Mpc larger than the TRGB distance to NGC\,1052--DF4 \citep[20.0 $\pm$ 1.6 Mpc,][]{Danieli2020} and this difference is statistically significant. Data for these two galaxies both come from \textit{HST}/ACS and are analyzed with the same methods, using the same forward modeling assumptions and the same absolute magnitude calibration.  The only difference between the two measurements is the depth of the data: in the luminosity function filter (F814W), there are 8 orbits for NGC\,1052--DF4 and 20 orbits for NGC\,1052--DF2. This translates into a difference in the amplitude of the photometric errors of $\approx 0.05$ mag.
We conservatively assume that the uncertainty in the relative TRGB magnitudes of the two galaxies is the quadratic sum of the forward modeling fit uncertainties and this 0.05\,mag difference. This leads to a distance difference of $\Delta D = 2.1 \pm 0.5$\,Mpc.
Figure \ref{fig:discussion} displays a visual summary of these findings.
The relative distance from NGC\,1052--DF2 to NGC\,1052--DF4 is kept fixed in the bottom panels.

As shown in the bottom panels of Figure\ \ref{fig:discussion}, the relative distance between NGC\,1052--DF2 and NGC\,1052--DF4 is significantly larger than the virial diameter of NGC\,1052, which means that at least one of the galaxies is not bound to the group.
The virial diameter is $780\pm 54$\,kpc \citep{Forbes2019}, \edit1{calculated from the NGC\,1052 halo mass assuming an NFW profile with concentration parameter $c=7.0$ and }the error comes from their reported velocity dispersion uncertainty.
NGC\,1052--DF2 is perhaps the most likely candidate for being unbound,
given its large radial velocity compared to other galaxies in the group  \citep{vanDokkum2018Nature}, and
the fact that its distance is only just consistent with that of NGC\,1052. 

Previously proposed formation scenarios for dark matter deficient galaxies fall broadly into two categories: formation within the group and tidal stripping. 
The first group of theories includes formation in the chaotic gas-rich environment of the assembling central galaxy \citep{vanDokkum2018Nature} and formation in a QSO outflow from the central black hole in NGC\,1052. In particular, it has been proposed that
a high-velocity collision between two gas-rich dwarfs occurring in the protogroup environment could explain
both the unusual GCs and the lack of dark matter in these galaxies
\citep{Silk2019}. The large relative distance and the large relative velocity
could be consistent with such scenarios.
The second category of scenarios
involves a progenitor falling into a Milky-Way sized halo, losing gas due to supernova feedback and ram pressure and losing dark matter by tidal stripping \citep{Nusser2020}.
Tidal stripping of dark matter is possible if the progenitor is on a tightly bound and quite radial orbit, has a cored density structure for the dark halo \citep{Ogiya2021},
and if the accretion happened early on, $z>1.5$. These scenarios are -- at least
at face value -- more difficult to reconcile with at least one of the galaxies being no longer
bound to the group.
\edit2{If NGC\,1052--DF2 and NGC\,1052--DF4 formed in isolation, the intense early feedback due to their GC systems could explain their dark matter deficiency \citep{Trujillo-Gomez2020,Trujillo-Gomez2021}.}

The relative distance also places a new constraint on the interpretation of NGC\,1052--DF2 and NGC\,1052--DF4 in the context of MOND (see \S\,\ref{sec:intro}).
\citet{vanDokkum2018Nature} argued that the low velocity dispersion of
NGC\,1052--DF2 could falsify MOND, but they neglected the external field effect (EFE).
\citet{Kroupa2018}, \citet{Haghi2019}, and \citet{Muller2019MOND} found that the observed
velocity dispersions of both NGC\,1052--DF2 and NGC\,1052--DF4 are
consistent with MOND if both
galaxies are physically close 
\citep[$<300$\,kpc; ][]{Kroupa2018}
to a massive galaxy.
The most obvious candidate for this massive galaxy is NGC\,1052 itself, but
our relative distance measurement of $\Delta D = 2.1 \pm 0.5$\,Mpc rules out
 NGC\,1052--DF2 and NGC\,1052--DF4 both being within 300\,kpc of NGC\,1052.
We cannot exclude that two different galaxies independently produce the EFE, as NGC\,1052--DF4 is close
in projection to the low luminosity spiral galaxy NGC1035 and it has been suggested
that it is interacting with it \citep{Montes2020}.
The Tully-Fisher distance of NGC1035 is consistent with it being in the foreground of the
group \citep[$15.6 \pm 2.2$\,Mpc, $v = 1392$\,km\,s$^{-1}$;][]{Kourkchi2017}. 
The issue with this scenario is that NGC\,1052--DF2 and NGC\,1052--DF4 would have formed completely
independently, one in a massive group and the other as a satellite of an isolated low luminosity
galaxy in the foreground, despite their near-identical morphologies, kinematics, and
extreme globular cluster populations.
Moreover, deep Dragonfly imaging does not confirm the bridge of stars between NGC\,1052--DF4 and NGC1035
that was found by \citet{Montes2020} \ (M.\ Keim et al., in preparation).


Looking ahead, further constraints on the physical processes that led to the formation of NGC\,1052--DF2 and NGC\,1052--DF4 can be obtained by combining the new distance information with
the GC luminosity functions \citep{Leigh2020,Trujillo-Gomez2020}, GC dynamics \citep[e.g. orbital decay timescale;][]{Nusser2018,DuttaChowdhury2019}, and tidal features \citep{Muller2019tidal}.
\deleted{Although \citet{Muller2019tidal} found no tidal features around NGC\,1052--DF2 and NGC\,1052--DF4, \citet{Montes2020} detected tidal tails in NGC\,1052--DF4 oriented towards NGC1035. We have obtained ultra deep Dragonfly images that
shed new light on these results (M.\ Keim et al., in preparation).}


\acknowledgments
We thank the referee for helpful comments that improved the manuscript. We thank Stacy McGaugh, Pavel Kroupa, Oliver M\"uller, Hosein Haghi, and Indranil Banik
for insightful discussions on MOND and we thank John Blakeslee for comments on SBF distances.
Support from STScI grants \textit{HST} GO-15851,14644, and 15695 are gratefully acknowledged.
ZS is supported by the Gruber Science Fellowship.
S.D. is supported by NASA through Hubble Fellowship grant HST-HF2-51454.001-A awarded by the Space Telescope Science Institute, which is operated by the Association of Universities for Research in Astronomy, Incorporated, under NASA contract NAS5-26555.
AJR is supported by National Science Foundation grant AST-1616710 and as a Research Corporation for Science Advancement Cottrell Scholar.
J.M.D.K.\ gratefully acknowledges funding from the Deutsche Forschungsgemeinschaft (DFG, German Research Foundation) through an Emmy Noether Research Group (grant number KR4801/1-1) and the DFG Sachbeihilfe (grant number KR4801/2-1), as well as from the European Research Council (ERC) under the European Union's Horizon 2020 research and innovation programme via the ERC Starting Grant MUSTANG (grant agreement number 714907).
%

\vspace{5mm}
\facilities{\textit{HST}(ACS)}


\software{astropy \citep{astropy2018}, 
DOLPHOT \citep{2000PASP..112.1383D}, 
Tiny Tim \citep{1995ASPC...77..349K}          }




\bibliography{paper.bib}{}
\bibliographystyle{aasjournal}



\end{document}